\documentstyle[preprint,eqsecnum,aps,prb]{revtex}
\begin{document}
\draft
\title{Anomalous effective charges and far IR optical absorption
of Al$_2$Ru from first principles}
\author{Serdar \"{O}\u{g}\"{u}t\cite{Serdar} and Karin M. Rabe}
\address{Department of Physics, Yale University\\
P. O. Box 208120, New Haven, Connecticut, 06520-8120}
\date{\today}
\maketitle
\begin{abstract}

For the orthorhombic intermetallic semiconductor Al$_2$Ru,  
the bandstructure, valence charge density, zone
center optical phonon frequencies, and Born effective charge
and electronic dielectric tensors are calculated using
variational density functional perturbation theory with
{\em ab initio} pseudopotentials and a plane wave basis set.
Good agreement is obtained with recent measurements on 
polycrystalline samples which showed 
anomalously strong far IR absorption by optical phonons, while
analysis of the valence charge density shows that the static ionic
charges of Al and Ru are negligible.
Hybridization is proposed as the single origin 
both of the semiconducting gap  
and the anomalous Born effective charges.
Analogous behavior is expected in related compounds such as
NiSnZr, PbTe, skutterudites, and Al-transition-metal quasicrystals.

\end{abstract}


Al$_2$Ru is an intermetallic semiconductor \cite{Evers} quite
close in composition to the Al-transition-metal quasicrystals.
Because of its relatively simple crystal structure, 
it has been suggested\cite{Manh,basov} that investigation of Al$_2$Ru
can contribute to the understanding of the pseudogap in
the electronic density of states observed in quasicrystals.
From first-principles calculations of the bandstructure and
charge density performed using linear-muffin-tin-orbitals (LMTO)
in the atomic sphere approximation,\cite{Manh}
it has been shown that the fundamental gap results from the
hybridization of $sp$(Al) and $d$(Ru) states. More recently, 
hybridization has been identified
as the origin of the pseudogap in other aluminum-rich 
transition-metal compounds.\cite{Manha,Manhb}

This high degree of hybridization in states near the fundamental gap 
might be expected to have implications for other properties as well.
In perovskite-structure oxides, it has been shown that
hybridization between oxygen $p$
orbitals and transition-metal $d$ orbitals results in values for
Born effective charges much larger than 
the nominal ionic values.\cite{Resta,Zhong,Ghosez}
In Al$_2$Ru, experimental measurement of the optical
conductivity shows a strikingly large oscillator
strength for phonon absorption.\cite{basov,basova} The estimated
values of Born effective charges, comparable to those of
typical ionic insulators, are very much
larger than the static ionic charges, which previous 
calculation\cite{Manh} has shown to be negligible.
In this paper, we report the first-principles 
calculation of the zone-center phonon frequencies, the Born
effective charges, and the dielectric tensor of Al$_2$Ru.
We obtain good agreement for the far-infrared optical conductivity
with the experimental measurements.
Confirming that the static charge transfer 
between Al and Ru is negligible, we conclude that hybridization 
may be responsible for the large effective charges, providing a
single mechanism 
linking the formation of the semiconducting gap and the
observed strong phonon optical absorption in the far IR.

The first principles calculations were performed using the
{\em ab initio} pseudopotential method with a plane wave basis
set and the conjugate gradients algorithm.\cite{Payne}
For Al, we used Hamann-Schl\"{u}ter-Chiang pseudopotentials\cite{HSC}
with reference configuration 3$s^1$3$p^1$3$d^1$ and cutoff radii of
1.1, 1.2, and 1.3 a.u. 
For Ru, we constructed scalar relativistic optimized pseudopotentials  
\cite{Rappe} with reference configuration 5$s^{1.75}$5$p^{0.25}$4$d^6$,
and cutoff radii of 2.2, 2.4, and 2.05 a.u. The pseudowavefunctions for 
$s$, $p$, and $d$ orbitals were expanded using 3, 4, and 6 Bessel
functions with $q_c$'s of 4, 4, and 6.05 $\sqrt{Ry}$, respectively. 
The pseudopotentials were put into separable form with two
projectors\cite{Proj} for each angular momentum with the $l=0$
component taken as local for both Al and Ru. 
The energy cutoff was taken as 500 eV, resulting 
in a kinetic energy convergence of 2 mRy. 
The local density approximation (LDA) was used with
the exchange-correlation functional of Ceperley and Alder
as parametrized by Vosko, Wilk and Nusair.\cite{Excorr}
The conjugate gradients
minimization for the self-consistent charge density and 
bandstructure was performed using the program {\sc castep} 2.1.
The force constants, Born effective charge tensors
$Z^*_{\kappa,\alpha\beta}$, and the dielectric
tensor $\epsilon_{\infty}$ were computed using the density functional 
perturbation theory method\cite{BGT} in the variational
formulation.\cite{GAT}
All calculations were performed for the experimentally
observed face-centered orthorhombic structure
with six atoms per unit cell (space group $Fddd$), 
lattice constants 
$a=8.015$ \AA, $b=4.715$ \AA, and $c=8.780$ \AA, and free
structural parameter $x=\frac{1}{3}$.\cite{structure}
The special $\bf{k}$-point set was a 4$\times$6$\times$4
Monkhorst-Pack grid\cite{Monkhorst} in the 
Brillouin zone (BZ) of a simple
orthorhombic lattice with lattice constants
$\frac{a}{2}$, $\frac{b}{2}$, and $\frac{c}{2}$, folded in to obtain
48 $\bf{k}$ points in the full BZ of the primitive face-centered 
orthorhombic lattice [Fig. 1(a)]. The acoustic sum rule 
for phonons and charge neutrality
for $Z^*_{\kappa,\alpha\beta}$ were imposed by adding
small corrections ($\leq 1\%$)
to the largest matrix elements in each case.

The first-principles bandstructure of Al$_2$Ru is shown in Fig. 1(b),
with the main features in good agreement with previously calculated  
bandstructures.
\cite{Manh,Burkov} At the bottom of the valence band
are free-electron-like bands derived mainly from 
Al $s$ states. In the upper half of the valence band, flat bands
derived from Ru $d$ states are seen to be strongly hybridized with
the more dispersive Al $s$ and $p$ bands. These calculations show that
Al$_2$Ru is a semiconductor, with a 0.35 eV indirect band gap 
from $\Gamma$ to the conduction band minimum along $\Gamma-B$.
The total valence charge density in the $z=0$ plane, shown in Fig. 2(a), is quite similar to the valence charge density of superimposed free neutral atoms. 
The difference, shown in Fig. 2(b), can be characterized as a small shift of charge from the core regions into the interstitial region between the Al and Ru atoms.
From this, it seems clear that there is 
no significant charge transfer between Al and Ru, and therefore, 
negligible ionic character for the bonding in this compound.

From our first-principles computations of the dielectric tensor,
we obtain
\begin{equation}
\epsilon_\infty= \left( \begin{array}{ccc}
18.9 & 0 & 0 \\
0 & 22.9 & 0 \\
0 & 0 & 20.7 \end{array}\right)
\end{equation}
These large values are characteristic of narrow-gap semiconductors,
such as PbTe, with $\epsilon_\infty=32.8$.\cite{pbte}
From computations of the Born effective charge tensors
$Z^*_{\kappa,\alpha\beta}=\partial P_{\alpha}/
\partial u_{\kappa,\beta}$, we find
for Ru, with site symmetry $222$,
\begin{equation}
Z^*_{Ru}=\left( \begin{array}{ccc}
-6.28 & 0 & 0 \\
0 & -6.96 & 0 \\
0 & 0 & -5.40 \end{array}\right)
\end{equation}
For Al, in the lower symmetry site $2..$, we find
\begin{equation}
Z^*_{Al}=\left( \begin{array}{ccc}
3.14 & 0 & 0 \\
0 & 3.48 & \pm 0.70\\
0 & \pm 0.47 & 2.70 \end{array}\right)
\end{equation}
where the plus/minus signs refer to the two different 
orientations of the nearest
neighbor shell for symmetry-related Al atoms.
In contrast to the negligible static ionic charges, 
these dynamical effective charges correspond in 
magnitude to a complete transfer of the fourteen valence electrons
per unit cell to Ru, leaving the Al ions with an effective
charge of $+3$. 
The phonon absorption is determined by these dynamical
charges, and not by the static ionic charges, accounting 
for the large oscillator strength observed in Al$_2$Ru.

To compare these results quantitatively with the measured optical 
conductivity, we also need the phonon frequencies and eigenvectors
at {\bf q} = {\bf 0}. The point group is $D_{2h}$, with eight 
irreducible representations (irreps) \cite{grptheory}. The dynamical 
matrix takes a block diagonal form, with a 6$\times$6 block
corresponding to pure displacements in the $\hat x$ direction
($A$ and $B_1$ irreps) and a 12$\times$12 block 
corresponding to displacements along
$\hat y$ and $\hat z$ ($B_2$ and $B_3$ irreps).
The calculated frequencies of the eighteen 
phonons are listed in Table I with
their symmetry labels. There are five optically active phonons in
Al$_2$Ru, with symmetry labels $B_{1u}$, $B_{2u}$ and $B_{3u}$.
 
First, we calculate the real part of the optical conductivity
$\sigma_1(\omega)=\frac{\omega}{4\pi}\epsilon_2(\omega)$, using
an oscillator model\cite{Bruesch}
\begin{equation}
\sigma_1(\omega)=\frac{e^2}{6\pi c V_0}\sum_j \frac{\gamma_j\omega^2}
{(\omega_j^2-\omega)^2+(\gamma_j\omega)^2}S_j
\end{equation}
where $S_j$ is the oscillator strength of the $j$th mode
\begin{equation}
S_j=\sum_{\alpha=x,y,z}(\sum_{\beta=x,y,z}\sum_\kappa
M_\kappa^{-\frac{1}{2}}Z^*_{\kappa,\alpha\beta}e_\beta(\kappa j))^2
\end{equation}
and $\kappa$ runs over the six atoms in the unit cell with 
volume $V_0$, $M_\kappa$
is the ion mass, and $\omega_j$, $\vec e(\kappa j)$, and $\gamma_j$ 
are the $j$th dynamical matrix eigenfrequency, eigenvector, 
and broadening, respectively.
Since we are comparing with results from a polycrystalline sample,
we sum over all modes neglecting the 
polarization index, and divide by 3
to average over all directions. The experimental measurement\cite{private}
at 300 K, shown in Fig. 3(a), includes four peaks.
We associate these with the five IR-active modes expected from symmetry by assuming that the intense peak at 265 cm$^{-1}$ consists of two unresolved modes.
The calculated frequencies thus represent underestimates of between 12\% and 18\% of the experimental frequencies, in parentheses: 133 cm$^{-1}$ (151 cm$^{-1}$), 216 cm$^{-1}$ and 248 cm$^{-1}$ (265 cm$^{-1}$), 279 cm$^{-1}$ (336 cm$^{-1}$), and 353 cm$^{-1}$ (405 cm$^{-1}$).
Values of the broadening $\gamma_1$, $\gamma_2$  and $\gamma_3$ for plotting the theoretical $\sigma_1(\omega)$ are chosen to reproduce the intensities of the lowest two experimental peaks, with $\gamma_4$ and $\gamma_5$ chosen to reproduce the qualitative shape of $\sigma_1(\omega)$ at higher frequencies (Fig. 3(b)).
While the lowest calculated peak is in excellent agreement with experiment, the splitting of the second and third calculated peaks is seen to be slightly too large. In addition, their combined oscillator strength, obtained from the calculated eigenvectors, is rather too small relative to the upper two first-principles peaks.
However, the calculated value for the integrated oscillator strength 
$8\int\sigma_1(\omega)d\omega$
is 4.74$\times$10$^{28}$s$^{-2}$, in excellent agreement
with the experimental value of 
4.94$\times$10$^{28}$s$^{-2}$.
Since this quantity is mainly determined 
by $Z^*_{\kappa,\alpha\beta}$ and 
is relatively insensitive to the phonon frequencies and
eigenvectors, we conclude that the large first-principles
values for the effective charges agree quite well with experiment.
 
Since the real part of the optical conductivity is independent of the
dielectric tensor $\epsilon_{\infty}$, 
an experimental value for comparison
with our calculations must be obtained from the imaginary part of the
optical conductivity
$\sigma_2(\omega)=-\frac{\omega}{4\pi}\epsilon_1(\omega)$.
We fit the measured $\sigma_2$($\omega$) below 100 cm$^{-1}$ to 
$c_1 \omega + c_2 \omega^2$.
Setting $c_1$ = $-\frac{1}{4\pi}\overline \epsilon_\infty 
-\frac{e^2}{3 V_0}{\sum_j}
\frac{S_j}{\omega_j^2}$, we obtain an experimental value 
of the polycrystalline average
$\overline \epsilon_\infty$ of 17,
which is 18\% smaller than the average of the calculated values
$\frac{1}{3}(\epsilon_{xx}+\epsilon_{yy}+\epsilon_{zz})=20.8$. 
This overestimate is typical of the results for this quantity
calculated using LDA.\cite{Overestimate}

Our first-principles calculations show that large Born effective
charges are an intrinsic feature of Al$_2$Ru, consistent with
experimental observation of strong phonon absorption.
However, these large charges are not the result of a static charge
transfer typical of ionic compounds. Instead, 
given that the key role of hybridization in opening the semiconducting
gap in Al$_2$Ru has already been established,\cite{Manh} we propose that
the hybridized nature of states near the gap is also responsible for the  
anomalous effective charges 
in Al$_2$Ru, in direct analogy to the
perovskite oxides.\cite{Resta,Zhong,Ghosez}
This could be more precisely formulated with a tightbinding analysis of the first-principles bandstructure\cite{Harrison}, though the necessary parametrization is complicated by the relatively low symmetry of Al$_2$Ru.
On a more positive note, we expect that similar behavior can be observed in other
semiconductors with narrow gaps opened by hybridization.
Indeed, experimental indications of large phonon absorption are
available for systems including NiSnZr,\cite{Aliev}
PbTe,\cite{Burstein} and skutterudites,\cite{Fleurial}
and theoretical support for the non-ionic nature of 
the effective charges in PbTe
and the other rocksalt IV-VI compounds has been obtained with the  
empirical pseudopotential method.\cite{Littlewood}
Building on first-principles investigations and 
tight-binding parametrizations of the bandstructures
of these systems,\cite{Ogut,Rabe,Singh}
a general understanding of the mechanisms
producing anomalous effective charges could be achieved.
With a local formulation transferable to nonperiodic systems, this
approach could eventually promote the interpretation of
optical conductivity measurements\cite{basova} to elucidate
the nature of vibrational modes in quasicrystals.

We thank D. N. Basov, T. Timusk, U. V. Waghmare, 
Ph. Ghosez, and D. Vanderbilt
for useful discussions and unpublished material.
We thank M. C. Payne and V. Milman for the use of {\sc castep} 2.1.
This work was supported by NSF Grant No.
DMR-9057442. In addition, K.M.R thanks D. Vanderbilt and the 
Department of Physics and Astronomy of Rutgers University for their
hospitality during the completion of this work, and acknowledges 
the support of the Clare Boothe Luce Fund and the Alfred
P. Sloan Foundation.

\begin{table}
\caption{The calculated phonon frequencies at {\bf q} = {\bf 0}, in
cm$^{-1}$, grouped according to the symmetry labels given in
Ref. \protect\onlinecite{grptheory}. The oscillator strengths $S_j$ for the optically active modes
are given in parentheses.}
\begin{tabular}{ccccccccc}
&$A_{g}$ &
$B_{1g}$ & $B_{2g}$ & $B_{3g}$ & $A_{u}$ & 
$B_{1u}$ & $B_{2u}$ & $B_{3u}$\\
\tableline
{}&334 & 193 & 176 & 223 & 342 & 0 & 0 & 0\\
{}&{}  & 225 & 231 & 344 & {}  & 216 (2.2) & 133 (0.84) & 248 (1.1)\\
{}&{}  & {}  & 402 & 448 & {}  & {}  & 279 (1.9) & 353 (0.67)\\
\end{tabular}
\end{table}

\begin{figure}
\caption{(a) Irreducible wedge of the face-centered 
orthorhombic Brillouin zone of 
Al$_2$Ru, shown embedded in the Brillouin zone of the simple 
orthorhombic lattice with lattice constants 
$\frac{a}{2}$, $\frac{b}{2}$, $\frac{c}{2}$ used 
in the construction of the {\bf k}-point 
sampling set. (b) Bandstructure of Al$_2$Ru along 
high-symmetry lines of the
face-centered orthorhombic Brillouin zone shown in (a).
The indirect gap of 0.35 eV is from the top of the valence band at
$\Gamma$ to the bottom of the conduction band along $\Gamma-B$.}
\end{figure}

\begin{figure}
\caption{(a) Total valence charge density of Al$_2$Ru in the $z=0$ plane.
The central Ru atom is surrounded by a hexagon of Al atoms. Contour
intervals are 1.3$\times$10$^{-4}$ electrons per unit cell. Counting from Al towards Ru, the contour labels increase from 1 to 32, then decrease again to zero at the Ru nucleus.
(b) Difference of the valence charge density (a) from that of superimposed free neutral atoms (Al $s^2p$ and Ru $d^7s^1$).
Contour intervals are 1/3 of those in (a), with dashed lines indicating negative contour values.  Counting outward from Al, the first few contour labels are -2, -2, -1, 0, 1 and 2.}
\end{figure}

\begin{figure}
\caption{The real part of the optical conductivity of Al$_2$Ru, in
($\Omega$ cm)$^{-1}$:
(a) experimental measurement at 300 K (Ref. \protect{\onlinecite{basova}}); (b) theoretical prediction.
In (b), a background of 50 ($\Omega$ cm)$^{-1}$ has been added and values of $\gamma_j$= 25, 9.0, 4.5, 65 and 50 $cm^{-1}$  
($j=1,...,5$) chosen as described in the text.}
\end{figure}

\end{document}